\documentstyle[aps,amssymb]{revtex}
\begin{document}
\title{On the spherical two-dimensional electron gas}
\date{October 27, 2001}
\author{J.\ Tempere$^{1,2}$, I. F. Silvera$^{1}$ and J. T. Devreese$^{2}$}
\address{$^{1}$ Lyman Laboratory of Physics, Harvard University, 14-A Oxford Street,\\
Cambridge MA 02139, USA.\\
$^{2}$ Dept. Natuurkunde, Universiteit Antwerpen (UIA), Universiteitsplein\\
1, B2610 Antwerpen, Belgium.}
\maketitle

\begin{abstract}
We investigate the many-body properties of a two-dimensional electron gas
constrained to the surface of a sphere, a system which is physically
realized in for example multielectron bubbles in liquid helium. A
second-quantization formalism, suited for the treatment of a spherical
two-dimensional electron gas (S2DEG), is introduced. Within this formalism,
the dielectric response properties of the S2DEG are derived, and we identify
both collective excitations and a spectrum of single-particle excitations.
We find that the single-particle excitations are constrained to a
well-defined region in the angular momentum - energy plane. The collective
excitations differ in two important aspects from those of a flat 2DEG: on a
sphere, the `spherical plasmons' have a discrete frequency spectrum and the
lowest frequency is nonzero.
\end{abstract}


\bigskip

\section{Introduction}

When a film of electrons on a flat helium surface reaches a critical
density, the helium surface becomes unstable \cite{GorkovJETPlet18} and
multielectron bubbles form \cite{VolodinJETPLet26}. These multielectron
bubbles are spherical cavities in the helium liquid, containing from a few
to several tens of millions of electrons. The bubble radius $R$ is
determined by balancing the Coulomb repulsion of the electrons in the bubble
with the surface tension of the helium : for $N=10000$ electrons, the
typical bubble radius is one micron, and scales as $N^{2/3}$ \cite
{ShikinJETP27}. Density functional calculations \cite
{SalomaaPRL47,ShungPRB45} indicate that the electrons inside the bubble are
not spread out homogeneously, but instead form a thin spherical layer with
thickness $\delta \ll R$ and radius $\approx R-\delta ,$ anchored to the
surface of the helium, with a binding energy of the order of several kelvin 
\cite{SalomaaPRL47,ColeRMP46}. The electrons are free to move in the
directions tangential to the spherical helium surface, so that in effect
they form a spherical two-dimensional electron gas (S2DEG).

Our present analysis of the spherical two-dimensional electron gas, albeit
motivated by the study of multielectron bubbles, is equally relevant from a
fundamental point of view: it is the logical next step to take after the
analysis of the flat two-dimensional electron gas, a topic that keeps
drawing renewed attention. Furthermore, S2DEGs also appear in doped
semiconductor particles if carriers accumulate in a surface layer \cite
{SasakiPRB40}, in charged droplets \cite{RayleighPHIL14}, and in fullerenes,
where the multipole excitation modes with angular momentum $l=1,2,3,4$ of
the S2DEG have recently been investigated \cite{InaokaSS273}.

In this paper, we introduce a second-quantization formalism suited for the
description of the S2DEG (in Sec. II). On the basis of this formalism, the
angular-momentum dependent dielectric function of the S2DEG is derived
within the RPA framework in Sec. III. Results are given for the
single-particle and collective excitations of the S2DEG and the static
structure factor dependence on the angular momentum quantum number in Sec.
IV.

\bigskip

\section{Hamiltonian of the S2DEG}

Decompositions in plane waves $\varphi _{{\bf k}}({\bf r})\propto e^{i{\bf k}%
.{\bf r}}$ are well suited to handle the many-body problem in flat space,
but are not appropriate to describe the many-body problem on a spherical
surface. The non-interacting single-particle wave functions suitable for the
description of the S2DEG are spherical harmonics, $\varphi _{l,m}(\Omega
)=Y_{l,m}(\Omega ),$ where $\Omega $ represents the spherical angles $\Omega
=\{\theta ,\phi \}$, and $l,m$ represent the angular momentum quantum
numbers. We introduce the creation operator $\hat{c}_{l,m}^{+}$ which
creates an electron in the state $\varphi _{l,m}(\Omega )$ on the sphere if
no electron is present in that state, and the annihilation operator $\hat{c}%
_{l,m}$ which destroys an electron in the single-particle state $\varphi
_{l,m}(\Omega )$.

\subsection{Kinetic energy}

Within second quantization theory (see e.g. \cite{Schweber}), the kinetic
energy operator can be written as a function of $\hat{c}_{l,m}^{+}$ and $%
\hat{c}_{l,m}$ as follows 
\begin{eqnarray}
\hat{T} &=&\sum_{l=0}^{\infty }\sum_{m=-l}^{l}\sum_{l^{\prime }=0}^{\infty
}\sum_{m^{\prime }=-l^{\prime }}^{l^{\prime }}\hat{c}_{l,m}^{+}\left[ \int
d\Omega \text{ }Y_{l,m}^{\ast }(\Omega )\frac{-\hbar ^{2}}{2m_{\text{e}}R^{2}%
}\Delta _{\Omega }Y_{l^{\prime },m^{\prime }}(\Omega )\right] \hat{c}%
_{l^{\prime },m^{\prime }}  \nonumber \\
&=&\sum_{l=0}^{\infty }\sum_{m=-l}^{l}\frac{\hbar ^{2}l(l+1)}{2m_{\text{e}%
}R^{2}}\hat{c}_{l,m}^{+}\hat{c}_{l,m}.  \label{Ekin}
\end{eqnarray}
In this expression, $m_{\text{e}}$ is the electron mass, $R$ is the radius
of the spherical electron gas, $\Delta _{\Omega }$ is the angle-dependent
part of the Lagrangian in spherical coordinates, and $\int d\Omega
=\int_{0}^{2\pi }d\phi \int_{0}^{\pi }d\theta $ $\sin \theta $. Expression (%
\ref{Ekin}) shows that the energy of a free electron in an angular momentum
state $\{l,m\}$ is 
\begin{equation}
E_{l,m}=\frac{\hbar ^{2}l(l+1)}{2m_{\text{e}}R^{2}}.
\end{equation}
If more than one electron is present, they will occupy the lowest-energy
states up to the Fermi energy $E_{F}$ characterized by a Fermi angular
momentum $L_{F}$, so that $N=2(L_{F}+1)^{2}$ and $E_{F}=\hbar
^{2}L_{F}(L_{F}+1)/(2m_{\text{e}}R^{2})$. Since in multielectron bubbles,
the bubble radius $R\propto N^{2/3}$ \cite{ShikinJETP33}, the Fermi energy
scales as $E_{F}\propto N^{-1/3}$ so that adding electrons to the
multielectron bubbles lowers the Fermi energy.

\subsection{Density}

The surface density on the sphere, defined as the expectation value of the
number of electrons in an infinitesimal spherical angle around $\Omega $, is 
\begin{equation}
\hat{n}_{\text{e}}(\Omega )=\sum_{l=0}^{\infty
}\sum_{m=-l}^{l}\sum_{l^{\prime }=0}^{\infty }\sum_{m^{\prime }=-l^{\prime
}}^{l^{\prime }}Y_{l,m}^{\ast }(\Omega )Y_{l^{\prime },m^{\prime }}(\Omega )%
\hat{c}_{l,m}^{+}\hat{c}_{l^{\prime },m^{\prime }}.  \label{ne}
\end{equation}
The decomposition in spherical harmonics of this density operator is given
by 
\begin{eqnarray}
\hat{n}_{\text{e}}(\Omega ) &=&\sum_{l=0}^{\infty }\sum_{m=-l}^{l}\hat{\rho}%
_{\text{e}}(l,m)Y_{l,m}^{\ast }(\Omega ),  \label{rho-0} \\
\hat{\rho}_{\text{e}}(l,m) &=&\int d\Omega \text{ }\hat{n}_{\text{e}}(\Omega
)Y_{l,m}(\Omega ).
\end{eqnarray}
Using the parity property of spherical harmonics, $Y_{l,m}(\Omega
)=(-1)^{m}Y_{l,-m}^{\ast }(\Omega )$, it is easily shown that $\hat{\rho}_{%
\text{e}}(l,m)=(-1)^{m}\hat{\rho}_{\text{e}}(l,-m)$. Substituting (\ref{ne})
in (\ref{rho-0}), we find for the spherical components of the density
operator: 
\begin{equation}
\hat{\rho}_{\text{e}}(l,m)=\sum_{l_{1}=0}^{\infty
}\sum_{m_{1}=-l_{1}}^{l_{1}}\sum_{l_{2}=0}^{\infty
}\sum_{m_{2}=-l_{2}}^{l_{2}}\left[ \int d\Omega \text{ }Y_{l,m}(\Omega
)Y_{l_{2},m_{2}}^{\ast }(\Omega )Y_{l_{1},m_{1}}(\Omega )\right] \hat{c}%
_{l_{2},m_{2}}^{+}\hat{c}_{l_{1},m_{1}}.  \label{6}
\end{equation}
The addition rules for spherical harmonics allow to perform the integral of
the product of spherical harmonics in (\ref{6}): 
\begin{eqnarray}
\hat{\rho}_{\text{e}}(l,m) &=&\sum_{l^{\prime }=0}^{\infty }\sum_{m^{\prime
}=-l^{\prime }}^{l^{\prime }}\sum_{L=|l-l^{\prime }|}^{l+l^{\prime
}}\sum_{M=-L}^{L}\sqrt{\frac{(2l+1)(2l^{\prime }+1)}{(2L+1)}}  \nonumber \\
&&\times \left\langle l,0;l^{\prime },0|L,0\right\rangle \left\langle
l,m;l^{\prime },m^{\prime }|L,M\right\rangle \hat{c}_{L,M}^{+}\hat{c}%
_{l^{\prime },m^{\prime }}.
\end{eqnarray}
In this expression, $\left\langle l,m;l^{\prime },m^{\prime
}|L,M\right\rangle $ is the Clebsch-Gordan coefficient for combining the
angular momenta $\{l,m\}$ and $\{l^{\prime },m^{\prime }\}$ into $\{L,M\}$.
To simplify this expression, we introduce the following notation: 
\begin{equation}
\hat{c}_{(l,m)\otimes (l^{\prime },m^{\prime })}^{+}=\sum_{L=|l-l^{\prime
}|}^{l+l^{\prime }}\sum_{M=-L}^{L}\sqrt{\frac{(2l+1)(2l^{\prime }+1)}{(2L+1)}%
}\left\langle l,0;l^{\prime },0|L,0\right\rangle \left\langle l,m;l^{\prime
},m^{\prime }|L,M\right\rangle \hat{c}_{L,M}^{+}.  \label{c-combo}
\end{equation}
The operator $\hat{c}_{(l,m)\otimes (l^{\prime },m^{\prime })}^{+}$ creates
an electron in a state which results from the combination of two angular
momentum states characterized by the quantum numbers $\{l,m\}$ and $%
\{l^{\prime },m^{\prime }\}$, respectively. With this notation, the
spherical components of the density operator can be written as 
\begin{equation}
\hat{\rho}_{\text{e}}(l,m)=\sum_{l^{\prime }=0}^{\infty }\sum_{m^{\prime
}=-l^{\prime }}^{l^{\prime }}\hat{c}_{(l,m)\otimes (l^{\prime },m^{\prime
})}^{+}\hat{c}_{l^{\prime },m^{\prime }},
\end{equation}
which highlights the analogy with the definition of the Fourier transformed
density operator of the flat two-dimensional electron gas (2DEG).

\subsection{Coulomb interaction on the sphere}

The Coulomb potential energy term for $N$ electrons, in the Hamiltonian is, 
\begin{equation}
\hat{H}_{\text{coulomb}}=\frac{1}{2}\sum_{j=1}^{N}\sum_{j^{\prime }\neq
j=1}^{N}\frac{e^{2}}{4\pi \varepsilon _{0}}\frac{1}{|{\bf \hat{r}}_{j}-{\bf 
\hat{r}}_{j^{\prime }}|},
\end{equation}
where ${\bf \hat{r}}_{j}$ represents the position operator of electron $j$
and $\varepsilon _{0}$ is the vacuum permittivity. The Coulomb potential on
a sphere with radius $R$ can be straightforwardly expanded in spherical
harmonics: 
\begin{equation}
\hat{H}_{\text{coulomb}}=\frac{1}{2}\sum_{j=1}^{N}\sum_{j^{\prime }\neq
j=1}^{N}\frac{e^{2}}{4\pi \varepsilon R}\sum_{l=0}^{\infty }\sum_{m=-l}^{l}%
\frac{4\pi }{2l+1}Y_{l,m}(\hat{\Omega}_{j})Y_{l,m}^{\ast }(\hat{\Omega}%
_{j^{\prime }}),
\end{equation}
with $\hat{\Omega}_{j}$ the (spherical angle) position operator of electron $%
j$ on the sphere. In second quantization this becomes 
\begin{equation}
\hat{H}_{\text{coulomb}}=\sum_{l_{1},m_{1}}\sum_{l_{2},m_{2}}%
%TCIMACRO{\tsum}%
%BeginExpansion
\mathop{\textstyle\sum}%
%EndExpansion
_{l,m}(-1)^{m}\text{ }v(l)\text{ }\hat{c}_{(l_{1},m_{1})\otimes (l,-m)}^{+}%
\hat{c}_{(l_{2},m_{2})\otimes (l,m)}^{+}\hat{c}_{l_{2},m_{2}}\hat{c}%
_{l_{1},m_{1}},
\end{equation}
with $\sum_{l,m}=\sum_{l=0}^{\infty }\sum_{m=-l}^{l}$ and 
\begin{equation}
v(l)=\frac{e^{2}}{2\varepsilon R}\frac{1}{2l+1}.  \label{v(L)}
\end{equation}
This term in the Hamiltonian describes the interaction process, shown in
Fig. 1. The initial state consists of two electrons in angular momentum
states $\{l_{1},m_{1}\}$ and $\{l_{2},m_{2}\}$. In the final state one
electron is in an angular momentum state resulting from adding $\{l,m\}$ to
its original angular momentum, and the other electron is in a state
resulting from subtracting $\{l,m\}$ from its original angular momentum.
Hence, the Coulomb interaction on a spherical surface can be described as
the exchange of a virtual photon with a given angular momentum $\{l,m\}$,
with an amplitude for this process of $v(l)$. The total Hamiltonian for the
S2DEG is 
\begin{eqnarray}
\hat{H} &=&\sum_{l=0}^{\infty }\sum_{m=-l}^{l}\frac{\hbar ^{2}l(l+1)}{2m_{%
\text{e}}R^{2}}\hat{c}_{l,m}^{+}\hat{c}_{l,m}  \nonumber \\
&&+\sum_{l_{1},m_{1}}\sum_{l_{2},m_{2}}%
%TCIMACRO{\tsum}%
%BeginExpansion
\mathop{\textstyle\sum}%
%EndExpansion
_{l,m}(-1)^{m}\text{ }v(l)\text{ }\hat{c}_{(l_{1},m_{1})\otimes (l,-m)}^{+}%
\hat{c}_{(l_{2},m_{2})\otimes (l,m)}^{+}\hat{c}_{l_{2},m_{2}}\hat{c}%
_{l_{1},m_{1}}  \label{H}
\end{eqnarray}

\section{Response properties of the S2DEG}

When an external field $V_{\text{ext}}(\Omega ;\omega )=\sum_{\ell ,m}V_{%
\text{ext}}(l,m;\omega )Y_{l,m}(\Omega )$ that couples to the electron
density is applied, an induced density $\rho _{\text{ind}}(\Omega ,\omega
)=\sum_{l,m}\rho _{\text{ind}}$($l,m;\omega )Y_{l,m}(\Omega )$ is generated.
Within linear response theory, the spherical components of this induced
density can be written as 
\begin{equation}
\rho _{\text{ind}}(l,m;\omega )=\frac{1}{\hbar }V_{\text{ext}}(l,m;\omega )%
{\cal D}_{R}(l,m;\omega ),
\end{equation}
where ${\cal D}_{R}(l,m;\omega )$ is the retarded density-density Green's
function \cite{Pines}. If the external potential $V_{\text{ext}}(l,m;\omega
) $ is a Coulomb potential created by an external charge $\rho _{\text{ext}}$%
, a dielectric function depending on the angular momentum can be introduced
to describe the screening of this external Coulomb potential: 
\begin{eqnarray}
\varepsilon (l,m;\omega ) &=&1-\frac{\rho _{\text{ind}}(l,m;\omega )}{\rho _{%
\text{ext}}(l,m;\omega )} \\
&\Rightarrow &\frac{1}{\varepsilon (l,m;\omega )}=1+\frac{v(l)}{\hbar }{\cal %
D}_{R}(l,m;\omega ).  \label{eps}
\end{eqnarray}
The retarded Green's function ${\cal D}_{R}(l,m;\omega )$ is derived from
the density-density Green's function 
\begin{equation}
{\cal D}(l,m;t)=-i\left\langle \Psi _{0}\left| {\cal T}\left\{
\sum_{l_{1},m_{1}}\hat{c}_{(l,m)\otimes (l_{1},m_{1})}^{+}(t)\hat{c}%
_{l_{1},m_{1}}(t)\sum_{l_{2},m_{2}}(-1)\hat{c}_{(l,-m)\otimes
(l_{2},m_{2})}^{+}\hat{c}_{l_{2},m_{2}}\right\} \right| \Psi
_{0}\right\rangle ,
\end{equation}
where the expectation value is taken with respect to the many-body ground
state $\left| \Psi _{0}\right\rangle $, ${\cal T}$ is the time-ordering
operator, and 
\begin{equation}
\hat{c}_{l,m}(t)=e^{i\hat{H}t/\hbar }\hat{c}_{l,m}e^{-i\hat{H}t/\hbar }.
\end{equation}
Within the second quantization formalism for the S2DEG, ${\cal D}%
_{R}(l,m;\omega )$ is evaluated with standard Green's function techniques.
In the resulting Feynman graphs (the `polarization bubble' graphs, \cite
{Mahan}), the wave numbers are replaced by angular momenta as illustrated in
Fig 1. To lowest order in the interaction amplitude $v(l)$, we find 
\begin{eqnarray}
{\cal D}_{R}^{\text{(0)}}(l,m;\omega ) &=&\sum_{l^{\prime },m^{\prime
}}\sum_{L=|l-l^{\prime }|}^{l+l^{\prime }}\frac{(2l+1)(2l^{\prime }+1)}{4\pi
(2L+1)}n(l^{\prime },m^{\prime })[1-n(L,m+m^{\prime })]  \nonumber \\
&&\times \left| \left\langle l,0;l^{\prime },0|L,0\right\rangle \right|
^{2}\left| \left\langle l,m;l^{\prime },m^{\prime }|L,m+m^{\prime
}\right\rangle \right| ^{2}  \nonumber \\
&&\times \left( \frac{1}{\omega +(E_{l^{\prime },m^{\prime
}}-E_{L,m+m^{\prime }})/\hbar +i\eta }-\frac{1}{\omega +(E_{L,m+m^{\prime
}}-E_{l^{\prime },m^{\prime }})/\hbar +i\eta }\right) ,  \label{DR0}
\end{eqnarray}
with $\eta $ an infinitesimal number, and $E_{l,m}$ the energy of a free
electron in angular momentum state $\{l,m\}$ on the sphere and 
\begin{equation}
n(l,m)=\left\langle \Psi _{0}\left| \hat{c}_{l,m}^{+}\hat{c}_{l,m}\right|
\Psi _{0}\right\rangle ,
\end{equation}
the occupation number of state $\{l,m\}$. This result, derived with the
formalism described in Sec. II, is in agreement with the result of Inaoka 
\cite{InaokaSS273} for the susceptibility of the S2DEG, which was used to
investigate the $l=1,2,3,4$ multipole modes of the S2DEG \cite{InaokaSS273}.
In Sec. IV we extend this investigation to include excitations with large
angular momentum $l$ (so that $l/L_{F}>1$), keeping the regime studied by
Inaoka as a limiting case. By investigating a much broader portion of the
excitation spectrum, we identify the nature of the excitations and expose
novel properties of those excitations.

The contribution of lowest-order in $v(l)$ to the retarded density-density
Green's function determines the Hartree-Fock approximation (HF) to the
dielectric function (\ref{eps}), given here for reference: 
\begin{equation}
\varepsilon _{\text{HF}}(l,m;\omega )=\left[ 1+\frac{v(l)}{\hbar }{\cal D}%
_{R}^{\text{(0)}}(l,m;\omega )\right] ^{-1}.  \label{eps-HF}
\end{equation}
It also permits the calculation of the random-phase approximation (RPA) to
the dielectric function, through a Dyson series for ${\cal D}_{R}^{\text{RPA}%
}:$ 
\begin{equation}
{\cal D}_{R}^{\text{RPA}}=\frac{{\cal D}_{R}^{\text{(0)}}}{1-[v(l)/\hbar ]%
{\cal D}_{R}^{\text{(0)}}},
\end{equation}
so that 
\begin{equation}
\varepsilon _{\text{RPA}}(l,m;\omega )=1-\frac{v(l)}{\hbar }{\cal D}_{R}^{%
\text{(0)}}(l,m;\omega ).  \label{eps-RPA}
\end{equation}
Note that the spin degree of freedom is not explicitly taken into account in
the expressions above. The interparticle potential and the kinetic energy
are considered to be spin-independent in the present treatment. The central
quantity, from which properties of the S2DEG are derived in the current
treatment, is the `polarization bubble' diagram leading to ${\cal D}_{R}^{%
\text{(0)}}.$ For spin-independent interaction potentials, the spin degree
of freedom only leads to a degeneracy prefactor ${\cal D}_{R}^{\text{(0)}}$:
the degeneracy of the unperturbed levels is doubled from $2l+1$ to $2(2l+1)$.

{\sl \bigskip }

\section{Results and discussion}

\subsection{Single-particle excitations}

In this section we will use units such that $m_{\text{e}}=\hbar =R=1$. The
dielectric function can be clarified by considering Plemelj's rule. The
imaginary part of the RPA dielectric function becomes 
\begin{eqnarray}
%TCIMACRO{\func{Im}}%
%BeginExpansion
\mathop{\rm Im}%
%EndExpansion
\varepsilon _{\text{RPA}}(l,m;\omega ) &=&\pi v(l)%
%TCIMACRO{\dsum}%
%BeginExpansion
\mathop{\displaystyle\sum}%
%EndExpansion
\limits_{l^{\prime }m^{\prime }}%
%TCIMACRO{\dsum}%
%BeginExpansion
\mathop{\displaystyle\sum}%
%EndExpansion
\limits_{L=|l-l^{\prime }|}^{l+l^{\prime }}%
%TCIMACRO{\dfrac{(2l+1)(2l^{\prime }+1)}{4\pi (2L+1)}}%
%BeginExpansion
{\displaystyle{(2l+1)(2l^{\prime }+1) \over 4\pi (2L+1)}}%
%EndExpansion
n(l^{\prime }m^{\prime })[1-n(L,m+m^{\prime })]  \label{Imeps} \\
&&\times \left| \left\langle l,0;l^{\prime },0|L,0\right\rangle \right|
^{2}\left| \left\langle l,m;l^{\prime },m^{\prime }|L,m+m^{\prime
}\right\rangle \right| ^{2}\delta \lbrack \omega -(E_{L,m+m^{\prime
}}-E_{l^{\prime },m^{\prime }})].  \nonumber
\end{eqnarray}
The dynamic structure factor of the S2DEG is related to the dielectric
function by 
\begin{eqnarray}
S(l,m{\bf ;}\omega ) &=&-\frac{1}{\pi v(l)}%
%TCIMACRO{\func{Im}}%
%BeginExpansion
\mathop{\rm Im}%
%EndExpansion
\left[ \frac{1}{\varepsilon (l,m;w)}\right] \\
&=&-\frac{1}{\pi v(l)}\frac{%
%TCIMACRO{\func{Im}}%
%BeginExpansion
\mathop{\rm Im}%
%EndExpansion
[\varepsilon (l,m{\bf ;}\omega )]}{\left\{ 
%TCIMACRO{\func{Re}}%
%BeginExpansion
\mathop{\rm Re}%
%EndExpansion
[\varepsilon (l,m{\bf ;}\omega )]\right\} ^{2}+\left\{ 
%TCIMACRO{\func{Im}}%
%BeginExpansion
\mathop{\rm Im}%
%EndExpansion
[\varepsilon (l,m{\bf ;}\omega )]\right\} ^{2}},  \label{sdyna}
\end{eqnarray}
and can be interpreted as the probability that an excitation with given
angular momentum quantum number $\{l,m\}$ and energy $\hbar \omega $ can be
created. $S(l,m{\bf ;}\omega )$ is a quantity accessible to experiment, in
particular scattering experiments. From expression (\ref{Imeps}) we conclude
that $%
%TCIMACRO{\func{Im}}%
%BeginExpansion
\mathop{\rm Im}%
%EndExpansion
\varepsilon _{\text{RPA}}(l,m;\omega )$ is zero unless adding the angular
momentum $\{l,m\}$ and the energy $\hbar \omega $ to the ground state of the
S2DEG can excite a single electron from an occupied state [$n(l^{\prime
},m^{\prime })=1$] into an unoccupied state [$n(L,m+m^{\prime })=0$]. These
are the single-particle excitations (called the ``Landau continuum'' in the
case of the flat 2DEG). Fig. 2 illustrates this concept in relation to the
S2DEG. Three regimes can be distinguished:

\begin{itemize}
\item  {\bf Case 1}: $l<L_{F}$. In the left panel of Fig. 2, we show the
case of angular momentum $\{l,m\}=\{2,1\}$ imparted on the system. The
filled disks represent the occupied states, the open squares are allowed
final states which can be reached by adding the angular momentum $\{l,m\}$
to the angular momentum of an electron in the Fermi sphere, and the open
circles are forbidden final states as they are occupied and excluded by
Fermi statistics. In this example the Fermi sea is filled up to $L_{F}=3$.
Not all final states are accessible, since to excite an electron it has to
go into a final state which is unoccupied (the open squares). This means in
practice that the electrons which can participate in creating an excitation
of angular momentum $\{l,m\}$ are the ones close to the Fermi level (in
fact, those from level $L_{F}-l$ up to $L_{F}$).

\item  {\bf Case 2:} $2L_{F}>l>L_{F}$. This is shown in the middle panel of
Fig. 2 depicting single-particle excitations which have angular momentum $%
\{4,-3\}$. Now all the electrons in the Fermi sea can participate, but not
all resulting final states are unoccupied.

\item  {\bf Case} {\bf 3: }$l>2L_{F}$. An example is shown in the right
graph of Fig. 2, for single-particle excitations of angular momentum $\{8,1\}
$. Now not only can all electrons participate in the process, but also all
possible final states are unoccupied and thus accessible.
\end{itemize}

\noindent The maximum energy difference between initial and final
single-particle states in Fig. 2 is 
\begin{eqnarray}
\omega _{\text{max}}(l) &=&%
%TCIMACRO{\dfrac{\hbar ^{2}}{2m_{\text{e}}R^{2}}}%
%BeginExpansion
{\displaystyle{\hbar ^{2} \over 2m_{\text{e}}R^{2}}}%
%EndExpansion
\left[ (L_{F}+l)(L_{F}+l+1)-(L_{F})(L_{F}+1)\right]  \nonumber \\
&=&%
%TCIMACRO{\dfrac{\hbar ^{2}}{2m_{\text{e}}R^{2}}}%
%BeginExpansion
{\displaystyle{\hbar ^{2} \over 2m_{\text{e}}R^{2}}}%
%EndExpansion
\left[ l^{2}+l(2L_{F}+1)\right] .  \label{wmax}
\end{eqnarray}
The smallest energy difference in case 3 ($l>2L_{F})$ is: 
\begin{eqnarray}
\omega _{\text{min}}(l) &=&%
%TCIMACRO{\dfrac{\hbar ^{2}}{2m_{\text{e}}R^{2}}}%
%BeginExpansion
{\displaystyle{\hbar ^{2} \over 2m_{\text{e}}R^{2}}}%
%EndExpansion
\left[ (l-L_{F})(l-L_{F}+1)-(L_{F})(L_{F}+1)\right]  \nonumber \\
&=&%
%TCIMACRO{\dfrac{\hbar ^{2}}{2m_{\text{e}}R^{2}}}%
%BeginExpansion
{\displaystyle{\hbar ^{2} \over 2m_{\text{e}}R^{2}}}%
%EndExpansion
\left[ l^{2}-l(2L_{F}-1)-2L_{F}\right] .  \label{wmin}
\end{eqnarray}
The two frequencies $\omega _{\text{min}}(l)$ and $\omega _{\text{max}}(l)$
demarcate a region in the frequency versus angular momentum plane in which
the single-particle excitations lie. Fig. 3 shows the location of the
excitations of the spherical two-dimensional electron gas in the angular
momentum $l$ vs. frequency $\omega $ plane. The $\{l,\omega \}$ values
corresponding to a single-particle excitation for an S2DEG\ with $L_{F}=10$
are shown in Fig. 3 as a markers within the limiting frequencies $\omega _{%
\text{min}}$ and $\omega _{\text{max}}$ of the Landau continuum region shown
by full curves.

\subsection{Plasmons in a spherical 2D electron gas}

In addition to the single-particle excitations in the S2DEG, collective
excitations are also possible. ``Collective excitations'' are defined as
poles of the density-density Green's function ${\cal D}$, whereas
single-particle excitations are defined as poles of the single-particle
Green's function \cite{Pines}. This means that collective excitations appear
when $\varepsilon (l,m;\omega )=0$. In the flat 2DEG, these collective
excitations are called plasmon modes{\sl . }We will use the same terminology
for the S2DEG (occasionally using `spherical plasmons' when a distinction is
needed).

For the collective mode, the imparted angular momentum is shared between all
the particles. This means that in a collective mode with angular momentum $%
\{l,m\}$ the entire spherical shell of electrons will oscillate with an
amplitude proportional to $Y_{l,m}(\Omega )$. The frequency of the
collective modes (the plasma frequency $\omega _{\text{pl}}$) will depend on
the angular momentum of the mode $\{l,m\}.$ The plasma frequency $\omega _{%
\text{pl}}(l,m)$ of the S2DEG\ can be found by solving 
\begin{equation}
\left\{ 
\begin{array}{l}
1-%
%TCIMACRO{\dfrac{e^{2}}{2\varepsilon \hbar R}}%
%BeginExpansion
{\displaystyle{e^{2} \over 2\varepsilon \hbar R}}%
%EndExpansion
%TCIMACRO{\dfrac{1}{2l+1}}%
%BeginExpansion
{\displaystyle{1 \over 2l+1}}%
%EndExpansion
%TCIMACRO{\func{Re}}%
%BeginExpansion
\mathop{\rm Re}%
%EndExpansion
\left[ {\cal D}_{R,0}(l,m;\omega _{\text{pl}})\right] =0 \\ 
%TCIMACRO{\func{Im}}%
%BeginExpansion
\mathop{\rm Im}%
%EndExpansion
\left[ {\cal D}_{R,0}(l,m;\omega _{\text{pl}})\right] =0
\end{array}
\right. .
\end{equation}
For the flat electron gas the last condition means that the plasma branch
lies outside the region of single-particle excitations discussed in the
previous subsection. In Fig. 3, the spherical plasmon excitations are shown
as full diamonds. Fig. 4 illustrates how the plasmon branch depends on the
number of electrons in the S2DEG. The radius of the S2DEG has been chosen
equal to the equilibrium radius of a multielectron bubble with a given
number of electrons \cite{ShikinJETP33}. The bubble radius is found by
balancing the helium surface tension and the Coulomb repulsion as discussed
in the introduction. Fig 5. illustrates how the plasmon branch depends on
the radius of the S2DEG, for a fixed number of electrons. We find that as
the radius is decreased (i.e. the multielectron bubble is compressed), the
plasmon branch lies relatively closer to the upper frequency of the
single-particle excitation region (as shown in the inset of Fig. 5).
However, the energy scale, set by $\hbar ^{2}/(m_{\text{e}}R^{2})$,
increases with decreasing radius so that the absolute value of the plasmon
frequencies increases with decreasing radius. The plasmon frequency is
independent of the projection of the angular momentum $m$. The plasmon
branch of the S2DEG strongly differs from the plasmon branch of 2DEGs in two
important aspects: it is discrete and the lowest accessible plasmon
frequency is larger than zero despite the acoustic nature of the plasmon
branch.

\subsection{Sum rule and spectral weights}

Sum rules prove to be a remarkably useful tool in the analysis of spectra,
both experimentally and theoretically. The zeroth moment sum rule for the
dynamic structure factor defines the static structure factor: $S(l,m):$%
\begin{equation}
S(l,m)=%
%TCIMACRO{\dint}%
%BeginExpansion
\displaystyle\int %
%EndExpansion
\limits_{0}^{\infty }S(l,m;\omega )d\omega =-\frac{1}{\pi v(l)}%
%TCIMACRO{\dint}%
%BeginExpansion
\displaystyle\int %
%EndExpansion
\limits_{0}^{\infty }%
%TCIMACRO{\func{Im}}%
%BeginExpansion
\mathop{\rm Im}%
%EndExpansion
\left[ \frac{1}{\varepsilon (l,m;\omega )}\right] d\omega .
\end{equation}
Since the dynamic structure factor only depends on the magnitude of the
angular momentum $l$ and not on $m$, this will also be true for the static
structure factor. Fig. 6 shows the static structure factor as a function of
angular momentum $l${\sl .} The open squares are the result in the
Hartree-Fock approximation [using $\varepsilon _{\text{HF}}$, expression (%
\ref{eps-HF})], and the full circles are the result in the RPA approximation
[using $\varepsilon _{\text{RPA}}$, expression (\ref{eps-RPA})]. As in the
case of the flat 2DEG, the HF structure factor $S_{\text{HF}}(l)$ is linear
for small $l/L_{F}$, whereas the RPA structure factor approaches zero more
rapidly with decreasing $l$. In the inset of Fig. 6, the first frequency
moment of the dynamic structure factor, 
\begin{equation}
\left\langle \omega (l)\right\rangle =%
%TCIMACRO{\dint}%
%BeginExpansion
\displaystyle\int %
%EndExpansion
\limits_{0}^{\infty }\omega S(l,m;\omega )d\omega ,  \label{ffsr}
\end{equation}
is shown, again for both HF and RPA approximations. Inaoka \cite{InaokaSS273}
derived a sum rule for the first frequency moment of the angular momentum
dependent dynamic structure factor. In units $\hbar =m_{\text{e}}=R$, this
is 
\begin{equation}
\left\langle \omega (l)\right\rangle =\frac{l(l+1)}{2}.  \label{sumrule}
\end{equation}
From Fig. 6 it is clear that the HF dynamic structure factor obeys this sum
rule. The RPA dynamic structure factor can be written as a sum of a
contribution from the plasmon mode and the contribution $S_{\text{cont}}$
from single-particle excitations: 
\begin{equation}
S_{\text{RPA}}(l;\omega )=A(l)\delta \lbrack \omega -\omega _{\text{pl}%
}(l)]+S_{\text{cont}}(l;\omega ),
\end{equation}
with $A(l)$ the spectral weight of the plasmon branch. The inset of Fig 6.
shows $\int \omega S_{\text{cont}}(l,\omega )d\omega $. From the deficit of
the RPA dynamic structure factor without the plasmon mode, shown in the
inset of Fig. 6, the spectral weight of the plasmon mode can be derived: 
\begin{equation}
A(l)\omega _{\text{pl}}(l)=\frac{l(l+1)}{2}-%
%TCIMACRO{\dint}%
%BeginExpansion
\displaystyle\int %
%EndExpansion
\limits_{0}^{\infty }\omega S_{\text{cont}}(l,\omega )d\omega .
\end{equation}
The spectral weight of the plasmon mode, along with the RPA dynamic
structure factor in the region of single-particle excitations, is shown in
Fig. 7 for $L_{F}=20$. The RPA dynamic structure factor of the S2DEG is
shown color-coded and as a function of the angular momentum $l$ and the
energy $\hbar w$. In white regions the RPA dynamic structure factor is zero.
In colored regions, the RPA dynamic structure factor differs from zero. This
figure is the complement of Fig. 4: whereas Fig. 4 shows the location and
type of the possible excitations of the S2DEG as a function of angular
momentum and energy, this figure shows their corresponding spectral weight
as expressed by the dynamic structure factor. At small angular momentum ($%
l/L_{F}<0.5$), the plasmon branch carries the most spectral weight, and the
region of single-particle excitations only has a small fraction of the total
spectral weight. Within the region of single-particle excitations, there is
a local maximum in the dynamic structure factor around $l=L_{\text{F}}$.

\subsection{Discussion}

The dynamic structure factor $S(l,m;\omega )$ is the spectral function of
the density-density Green's function and as such can be interpreted as a
probability function (see e.g. \cite{Mahan}, p. 153). $S(l,m;\omega )d\omega 
$ is the probability that an excitation of the S2DEG has angular momentum
quantum numbers $\{l,m\}$ and energy between $\hbar \omega $ and $\hbar
(\omega +d\omega )$. This is analogous to the interpretation of the more
familiar dynamic structure factor [$S({\bf k},\omega )$] expressed as a
function of wave number, which is the probability that an excitation has
momentum $\hbar {\bf k}$ and energy $\hbar \omega $.

A `natural' experiment to determine the dynamic structure factor directly is
an inelastic scattering experiment. In particular the differential cross
section for a probe particle of mass $M$ (such as an incident ion) and
initial momentum ${\bf p}_{\text{i}}$ to be scattered to a final state ${\bf %
p}_{\text{f}}$ with energy transfer between $\hbar \omega $ and $\hbar
(\omega +d\omega )$ to the scattering system is \cite{Pines}: 
\begin{equation}
\frac{d\sigma }{d\theta d\omega }=\frac{M^{2}}{8\pi ^{3}}\frac{p_{f}}{p_{i}}%
|V_{k}|^{2}S(k,\omega ).
\end{equation}
In this expression, $\theta $ is the scattering angle between ${\bf p}_{%
\text{f}}$ and ${\bf p}_{\text{i}}$, $|V_{{\bf k}}|$ is the Fourier
transform of the interaction potential between the probe particle, and the
energy transfer $\omega $ is related to the momentum transfer ${\bf k}$ by $%
\hbar \omega =\hbar ^{2}{\bf k}.{\bf p}_{\text{i}}/m+(\hbar k)^{2}/(2m)$.
The relation between the Fourier decomposition and the spherical
decomposition of the density is

\begin{eqnarray}
\widehat{\rho }_{\text{e}}({\bf k}) &=&\sum_{j=1}^{N}\left\langle e^{i{\bf k}%
.\widehat{{\bf r}}_{j}}\right\rangle =\sum_{j=1}^{N}\sum_{l}\sqrt{4\pi (2l+1)%
}i^{l}j_{l}(kR)\left\langle Y_{l,0}(\widehat{\theta }_{j})\right\rangle 
\nonumber \\
&=&\sum_{l}\sqrt{4\pi (2l+1)}i^{l}j_{l}(kR)\widehat{\rho }_{\text{e}}(l,0)
\label{rhochange}
\end{eqnarray}
where the $z$-axis has been taken along the direction of the wave vector $%
{\bf k}$, and $j_{l}(x)$ is the spherical Bessel function of the first kind,
and $\theta _{j}$ is the angle between ${\bf k}$ and the position operator
of electron $j$, ${\bf r}_{j}$. The dynamic structure factors are related by 
\begin{equation}
S(k,\omega )=\sum_{l}\left| \sqrt{4\pi (2l+1)}j_{l}(kR)\right|
^{2}S(l,0;\omega )
\end{equation}
which can be derived by writing the dynamic structure factor as a
density-density correlation function in the Lehmann representation and
substituting expression (\ref{rhochange}) for $\widehat{\rho }_{\text{e}}(%
{\bf k})$.

Note that the dependence of the angular wave functions on the polar angle $%
\theta $\ becomes semiclassical when the quantum number $l$\ is large. In
particular, the WKB expression for the angular function becomes \cite{landau}
\begin{equation}
Y_{l0}(\theta )\approx \frac{i^{l}}{\pi }\frac{\sin [(l+1/2)\theta +\pi /4]}{%
\sqrt{\sin \theta }}  \label{semicl}
\end{equation}
This semiclassical approximation will be valid in the region where $\theta
l\gg 1$\ and $(\pi -\theta )l\gg 1$, which holds at large $l$\ for almost
all values of $\theta $\ except those close to the `poles' $\theta =0$\ and $%
\theta =\pi .$\ The semiclassical approximation (\ref{semicl}) corresponds
to the limit of a plane wave, where the wave length is given by $R(2\pi /l)$%
. As such, this approximation leads primarily to the results of a flat 2D
electron gas where $\hbar l/R$\ plays the role of the electron momentum.
This is valid in the limit that $R$\ is large and the ratio $l/R$\ remains
finite, although even for small $R$, this approximation will still hold at
large $l$. However, it does not reproduce some of the interesting low
angular momentum results such as a lower boundary to the plasmon
frequencies. 

\section{Conclusions}

In this paper, we have studied the properties of the two-dimensional
electron gas in a distinct geometry which is recently gaining interest in
connection to multielectron bubbles, namely electrons confined to a
spherical surface. For this purpose, we set up a second-quantization
formulation based on the spherical harmonics as single-electron building
blocks for the many-electron theory, introduced in Sec. II. Within this
formalism, the dynamic structure factor is derived in the RPA framework and
both the single-particle excitations and collective excitations are analyzed
for S2DEGs with $L_{F}$ up to $40$ and for angular momenta $l$ up to $%
3L_{F}. $

The dynamic structure factor in the RPA approximation reveals collective
excitations (`spherical plasmon modes'), which differ from the collective
modes of the flat 2DEG in that the spherical plasmon modes of the S2DEG are
discrete in frequency, and the smallest spherical plasmon frequency is
larger than zero. For $l/L_{F}<0.5$, we find that the spherical plasmon mode
carries the main fraction of the spectral weight. The single-particle
excitations are confined to a region determined by $\omega _{\text{min}}(l)$
and $\omega _{\text{max}}(l)$ given by expressions (\ref{wmin}) and (\ref
{wmax}). In the semiclassical approximation, the role of the electron
momentum $\hbar k$ in the flat 2DEG is played by $\hbar l/R$ in the
spherical electron gas, and the energy scale is set by $\hbar ^{2}/(m_{\text{%
e}}R^{2})$, which opens interesting prospects for multielectron bubbles
since the radius can be varied by either changing the number of electrons or
-independently- the pressure. For typical multielectron bubbles with $%
N\leqslant 10^{4}$, the plasmon modes of the S2DEG lie in the far-infrared,
and their frequency increases with decreasing number of electrons in the
multielectron bubble or with decreasing radius of the bubble, such that
these novel collective modes may be detectable in forthcoming experiments on
stabilized multielectron bubbles \cite{SilveraBAPS46}.

\section*{Acknowledgments}

Discussions with V. Fomin, S. Klimin and J. Huang are gratefully
acknowledged. J. Tempere is supported financially by the FWO-Flanders with a
mandate `Postdoctoral Fellow of the Fund for Scientific Research - Flanders'
(Postdoctoraal Onderzoeker van het Fonds voor Wetenschappelijk Onderzoek -
Vlaanderen). This research has been supported by the Department of Energy,
grant DE-FG002-85ER45190, and by the\ GOA BOF UA 2000, IUAP, the FWO-V
projects Nos. G.0071.98, G.0306.00, G.0274.01, WOG WO.025.99N (Belgium).

\bigskip

\newpage

\begin{center}
{\bf FIGURE\ CAPTIONS}
\end{center}

\bigskip

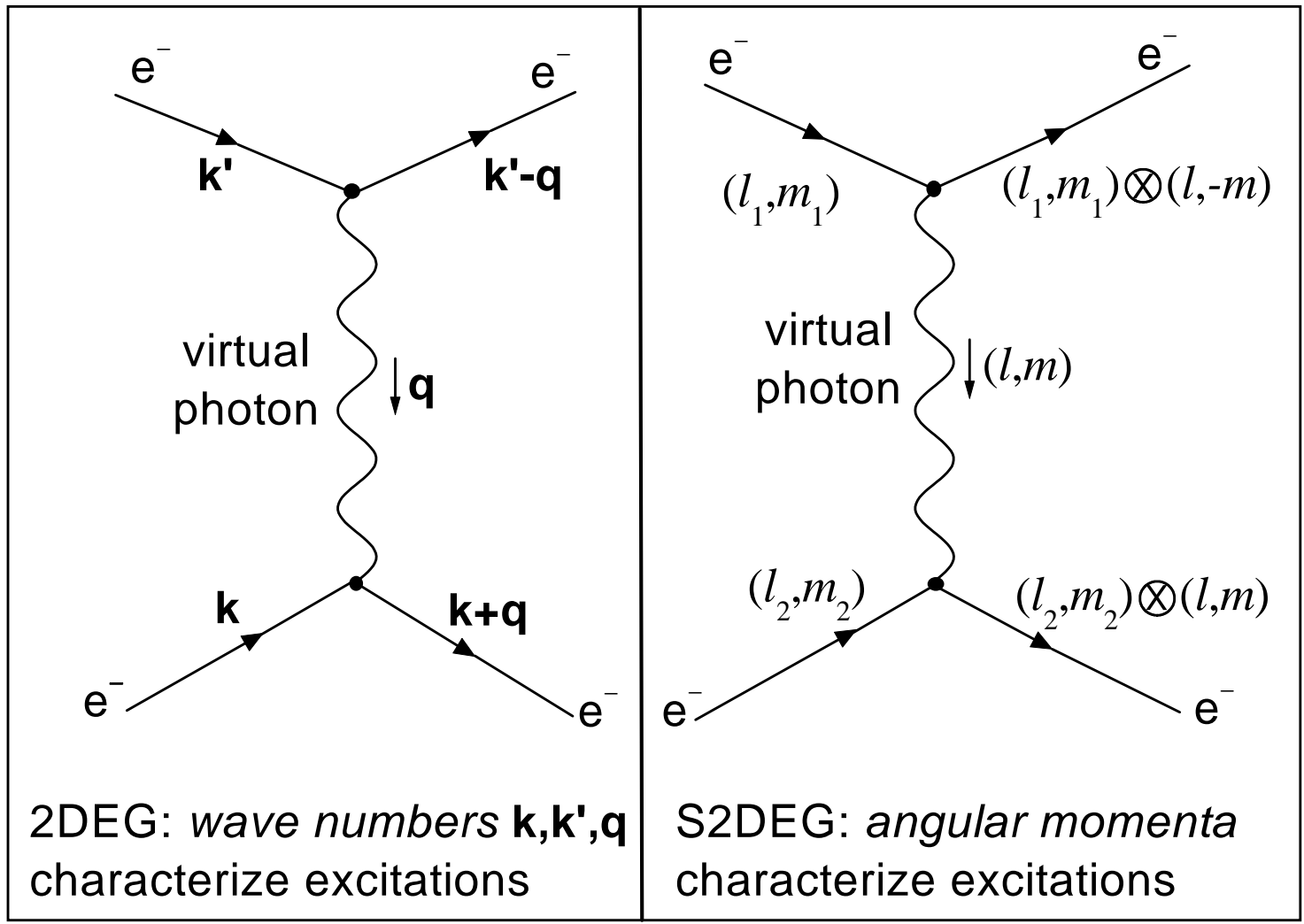
FIG. 1. This figure shows two Feynman diagrams for equivalent processes,
namely the exchange of a virtual photon between two electrons. The panel at
the left shows the diagram for the case of a flat 2DEG and the one to the
right is for the case of a S2DEG. In the flat 2DEG, wave numbers are good
quantum numbers, in the S2DEG they have to be replaced by angular momentum
quantum numbers. Vector addition of momenta has to be replaced by addition
of angular momenta.

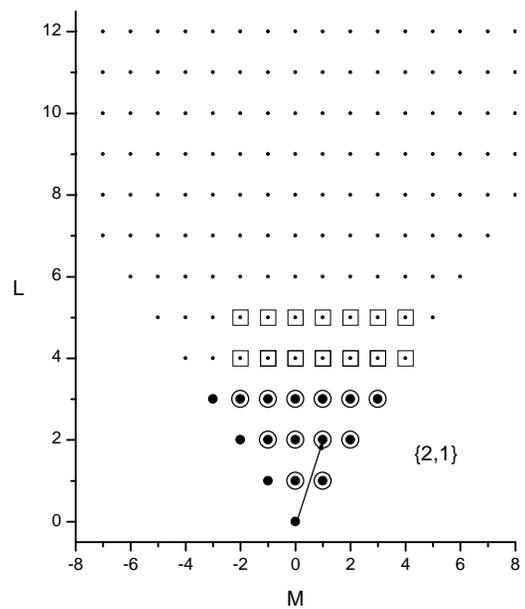
FIG 2. This figure shows diagrams representing the angular momentum states
of an electron on a spherical surface. In the three panels, electrons are
occupying the single-particle states up to a Fermi-level angular momentum $%
L_{F}=3$. This Fermi sea of occupied states is shown as filled circles. The
diagrams also show the final states which can occur from combining the
angular momentum of any of the electrons in the Fermi sea with the angular
momentum of an excitation (with $\{l,m\}=\{2,1\},\{4,-3\},\{8,1\}$ for the
left, middle and right panels, respectively). The final states which are
unoccupied, and thus allowed, are indicated by hollow squares. The final
states which cannot be achieved because there is already an electron present
are indicated by hollow circles.

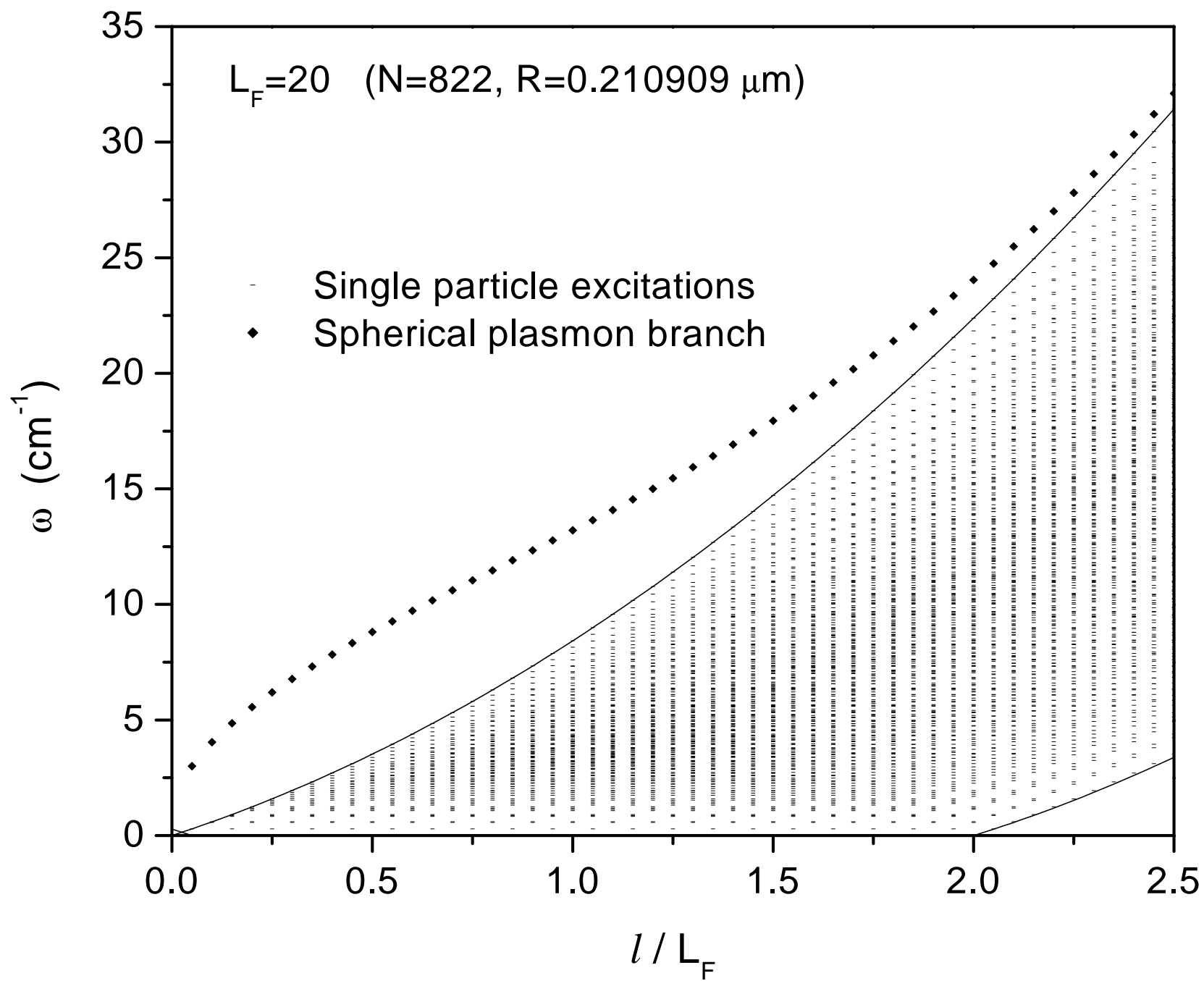
FIG. 3. In the many-body S2DEG, two types of excitations can be
distinguished: single particle excitations (which appear for $%
%TCIMACRO{\func{Im}}%
%BeginExpansion
\mathop{\rm Im}%
%EndExpansion
[\varepsilon ]\neq 0$) and collective excitations (which appear when $%
%TCIMACRO{\func{Re}}%
%BeginExpansion
\mathop{\rm Re}%
%EndExpansion
[\varepsilon ]=%
%TCIMACRO{\func{Im}}%
%BeginExpansion
\mathop{\rm Im}%
%EndExpansion
[\varepsilon ]=0$). The excitations of the S2DEG can be characterized by
their angular momentum $l$ and energy $\hbar \omega $. This figure shows the
location of the excitations in the frequency (energy) vs. angular momentum
plane. The single particle excitations are discrete, and all lie in a region
demarcated by the black curves, representing $\omega _{\text{min}}(l)$ and $%
\omega _{\text{max}}(l)$ given by expressions (\ref{wmin}),(\ref{wmax}). The
collective excitations are shown as filled diamonds, lying above the $\omega
_{\text{max}}(l)$ curve.

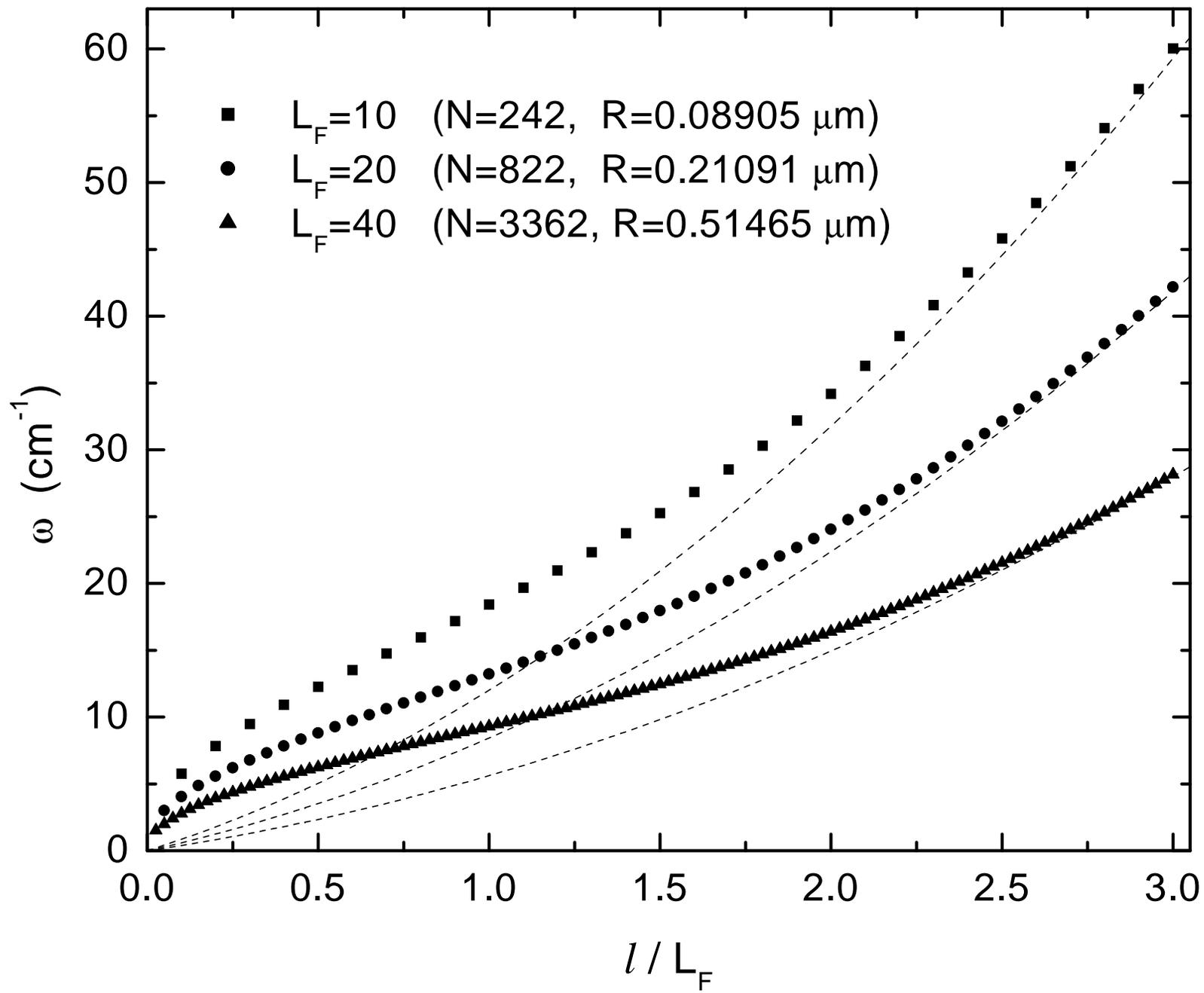
FIG. 4. The collective excitations of the S2DEG are the analogue of plasmons
in the flat 2DEG and appear at frequencies $\omega $ and angular momenta $l$
such that $%
%TCIMACRO{\func{Re}}%
%BeginExpansion
\mathop{\rm Re}%
%EndExpansion
[\varepsilon (l,\omega )]=%
%TCIMACRO{\func{Im}}%
%BeginExpansion
\mathop{\rm Im}%
%EndExpansion
[\varepsilon (l,\omega )]=0$. For S2DEGs with $L_{F}=10,20,40$, these
collective excitations are shown as filled squares, circles and triangles,
respectively. The dashed curves denote the upper frequency of the region of
single-particle excitations (given by expression (\ref{wmax})). As more
electrons are put into the multielectron bubble, the frequency of the
collective excitations decrease. As the Fermi level is increased, $%
L_{F}\rightarrow \infty $, the spherical plasmon excitations form a
continuous curve and the lowest spherical plasmon frequency $\omega _{\text{%
pl}}(l=1)$ approaches zero.

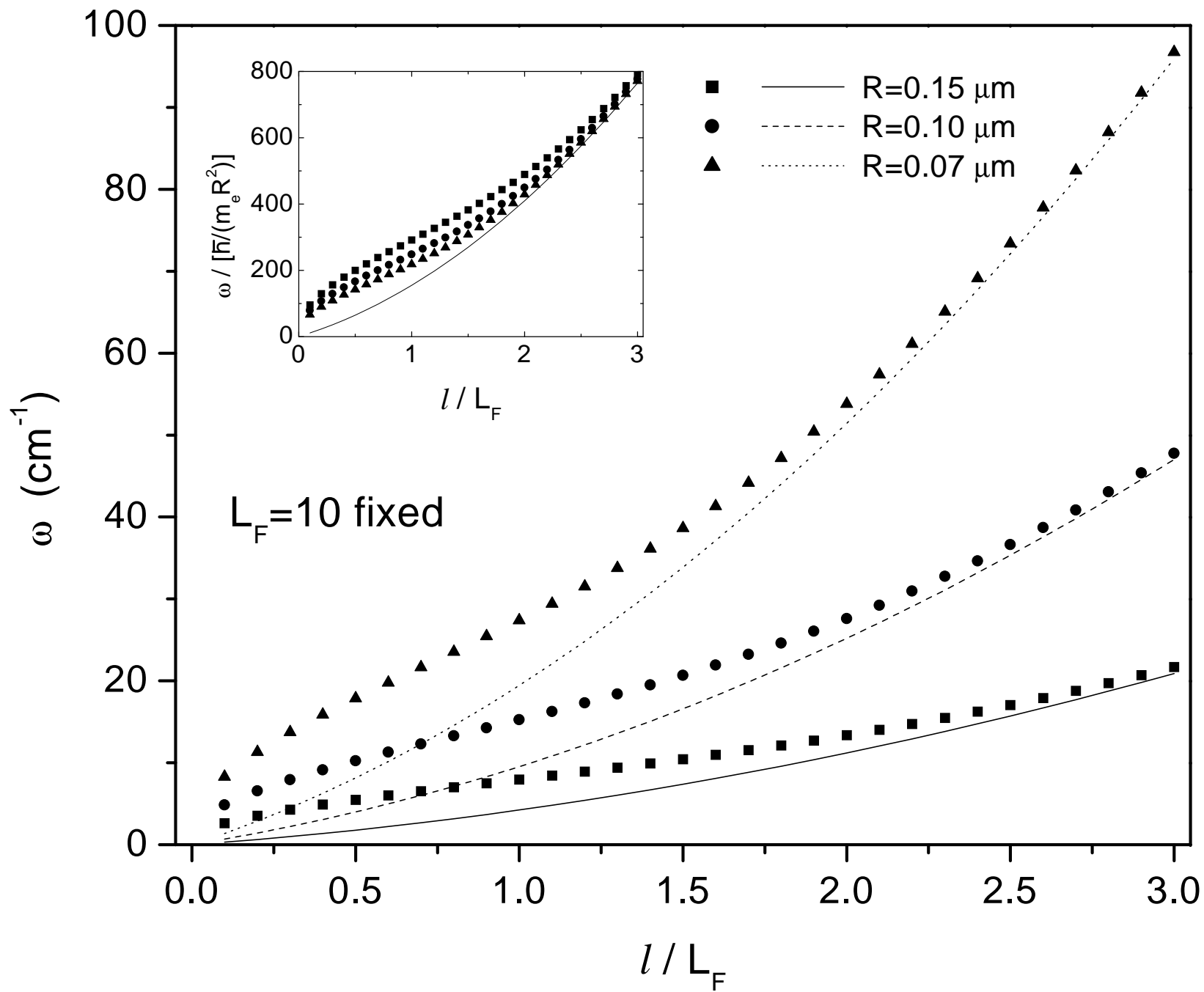
FIG. 5. In a multielectron bubble (MEB), the radius of the spherical
electron gas can be varied independently of the number of electrons by
applying pressure. This figure, which complements Fig. 4, shows the effect
of changing the MEB radius on the frequencies of the collective excitations
(the 'spherical plasmons') as a function of the angular momentum. The inset
shows the same results, with the frequency rescaled to the natural bubble
frequency $\hbar /(m_{\text{e}}R^{2})$.

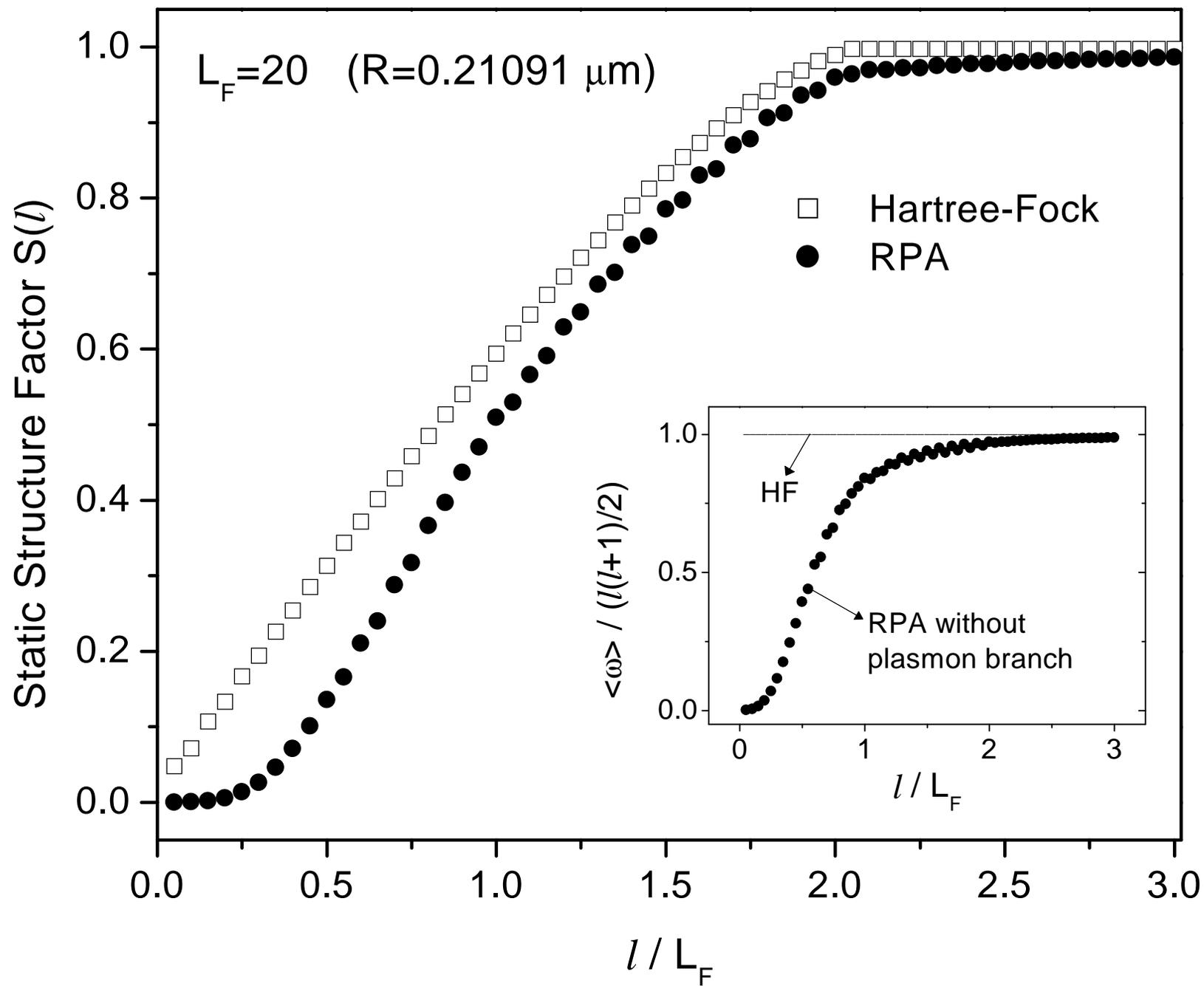
FIG. 6. The static structure factor $S(l)$ for a spherical two-dimensional
electron gas (S2DEG) with $L_{F}=20$ is shown as a function of $l$, both in
the RPA and in the Hartree-Fock approximations. Note that in the
Hartree-Fock approximation, $S(l)$ is linear in $l$ at small $l$, whereas in
the RPA approximation it goes to zero faster than linearly. In the inset,
the sum rule (\ref{sumrule}) for the first frequency moment of the dynamic
structure factor of the S2DEG \cite{InaokaSS273} is checked. The
Hartree-Fock result complies perfectly to the sum rule. The RPA\ result
without the plasmon branch does not have enough spectral weight to satisfy
the sum rule for $l/L_{F}<2$, which indicates that in this region a
substantial part of the spectral weight lies with the plasmon mode -- the
strength of the plasmon branch is derived from this result.

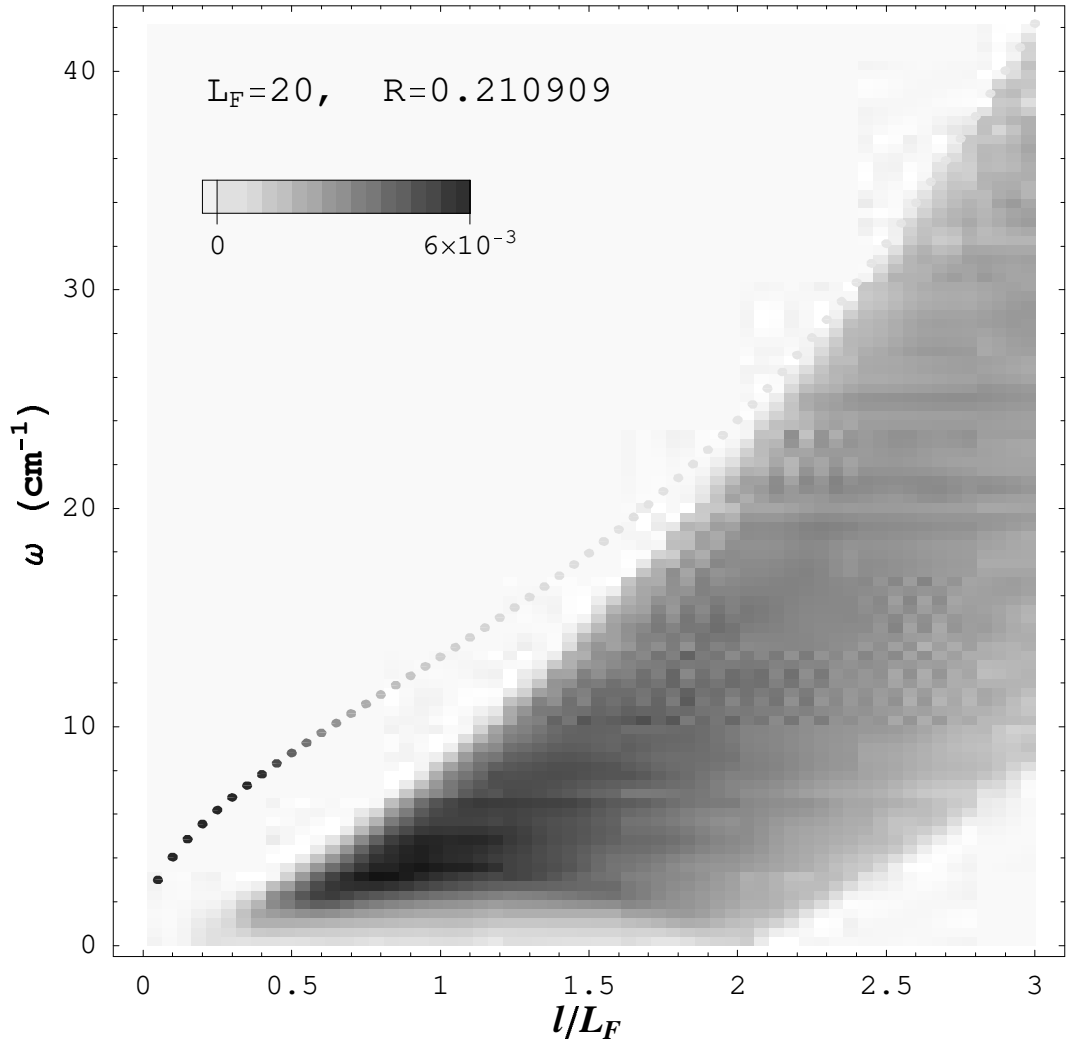
FIG. 7. The RPA dynamic structure factor $S(l,\omega )$ for a spherical
two-dimensional electron gas (S2DEG) with $L_{F}=20$ is shown here in units $%
\hbar =m_{\text{e}}=R=1$ as a function of the angular momentum quantum
number $l$ and their energy $\omega .$ At zero temperature, the RPA dynamic
structure factor of the S2DEG consists of a region of single-particle
excitations and a set of collective modes (colored disks). This figure
complements Fig. 3, which shows only the locations of the excitations,
whereas this figure shows their spectral weights.

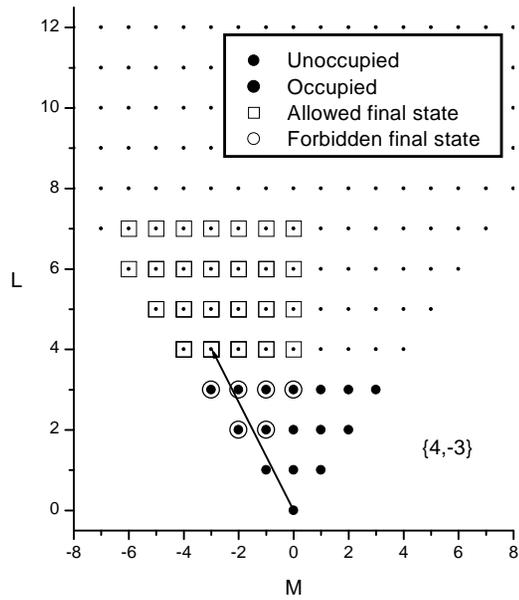
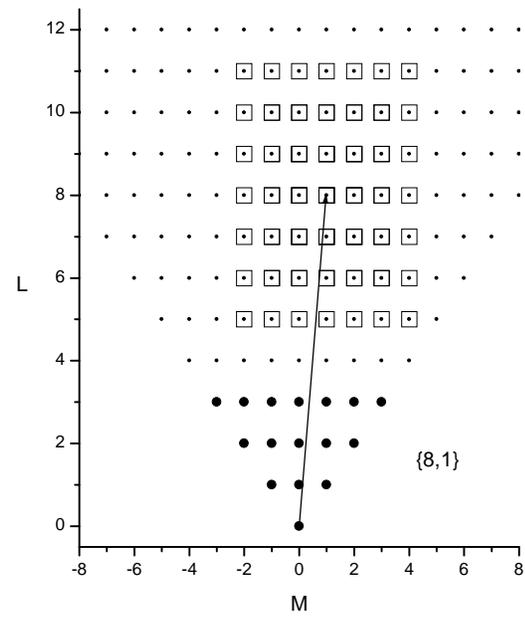
\end{document}